# Orbital Reconstruction in a Self-assembled Oxygen Vacancy Nanostructure

H. Jang[1], G. Kerr[1], J. S. Lim[2], C.-H. Yang[2], C.-C. Kao[3], and J.-S. Lee[1,*]

[1]Stanford Synchrotron Radiation Lightsource, SLAC National Accelerator Laboratory, Menlo Park, California 94025, USA
[2]Department of Physics, KAIST, Yuseong-gu, Daejeon 305-701, South Korea
[3]SLAC National Accelerator Laboratory, Menlo Park, California 94025, USA

*jslee@slac.stanford.edu

## ABSTRACT

We demonstrate the microscopic role of oxygen vacancies spatially confined within nanometer inter-spacing (about 1 nm) in $BiFeO_3$, using resonant soft X-ray scattering techniques and soft X-ray spectroscopy measurements. Such vacancy confinements and total number of vacancy are controlled by substitution of $Ca^{2+}$ for $Bi^{3+}$ cation. We found that by increasing the substitution, the in-plane orbital bands of $Fe^{3+}$ cations are reconstructed without any redox reaction. It leads to a reduction of the hopping between Fe atoms, forming a localized valence band, in particular Fe $3d$-electronic structure, around the Fermi level. This band localization causes to decrease the conductivity of the doped $BiFeO_3$ system.

# Introduction

In an intrinsic manner, oxygen vacancies always reside in all oxide compounds modulating chemical and physical properties on the scheme of defect chemistry. Beyond regarding this as an intrinsic defect, nowadays the oxygen vacancy has been considered as a parameter for controlling functionalities of oxide compounds such as quantum materials with strong electron correlation[1-5] and energy materials [6-9]. In this context, it has been demonstrated that changes in electronic conductivity on the correlated perovskite $AB$O$_3$[3,4] and Li-based batteries[6,8] are associated with total oxygen vacancies. Furthermore, the spatially inhomogeneous distribution of the oxygen vacancies is regarded as another important parameter, as demonstrated in multiferroic BiFeO$_3$[10-14]. Ferroelectric polarization in BiFeO$_3$ can be tuned by an external electric field. Meanwhile, the external electric field additionally induces oxygen vacancy migration because oxygen vacancies are positively charged[11,15], leading to a switchable photovoltaic effect in BiFeO$_3$[10-14,16]. In spite of the vacancy's importance in those applications, however, a role of oxygen vacancies has been puzzling and discussed only conceptually. Thus, the lack of microscopic understanding limits the improvement of the functionalities of oxide compounds.

For this reason, we discuss the microscopic role of oxygen vacancies for a representative multiferroic photovoltaic system, BiFeO$_3$. In general, the photovoltaic effect is associated with a modification of the electrons present in both the valence and conduction band when a material absorbs energy via the light[17]. However, the multiferroic BiFeO$_3$ case is more complicated because the oxygen vacancy's migration is also affected by the built-in electric field. Here, we investigate an electronic structure with varying the oxygen vacancy in $BiFeO_3$ using resonant soft X-ray scattering and soft X-ray absorption spectroscopy measurement and corresponding atomic model calculations. Considering previous works, the number of the oxygen vacancies in $BiFeO_3$ can be readily controlled with $Ca^{2+}$ substitution ($x$) for $Bi^{3+}$ cations – $Bi_{1-x}Ca_xFeO_{3-\delta}$ (hereafter, BCFO)[15,18]. Each planar defect an arrangement of Bi/Ca cations is adopted and promotes the formation of oxygen vacancy, showing a brownmillerite-like intra-plane, which leads to a superstructure (see Fig. 1a)[15,19]. Considering the previous TEM studies[19-21], furthermore, the oxygen vacancy in the superstructure is confined within a single unit cell in a self-assembled manner and the planar structures periodically appear at a few nanometers interval depending on the Ca substitution ratio (see Methods).

In the doped BCFO case, anionic electron number is reduced by the formation of positively charged oxygen-vacancy[15]. In this manner, electron-hole pairs are modified by the oxygen vacancy. According to the reported photovoltaic properties of BCFO as a function of the vacancy concentration[22], however, it does not show a monotonic increase even in a monotonic enhancement of the oxygen vacancy in BCFO. In particular, such effect decreases beyond $x \sim 15\%$[22]. This means that the reported diode effect[10,11] of $BiFeO_3$ cannot be simply employed for explaining a change in BCFO via the oxygen vacancy. This implies that near cations (i.e., Fe in this case) chemically responds to the oxygen vacancies, leading to our attention for a role of oxygen vacancies via spectroscopic scheme such as electronic configuration.

## Results

Figure 2a shows O K-edge X-ray absorption spectroscopy (XAS) spectra on $Ca^{2+}$ 25% doped BCFO sample (BCFO25), aiming to address the spectroscopy scheme of the oxygen vacancy. The spectra were acquired by recording the total electron yield (TEY) – details of the experimental geometry are shown in Fig. 2b (see Methods). The spectral features represent hybridization effects between O 2p-Fe 3d bands[23]. There are two pronounced features around $E$ = 527.4 eV and 528.8 eV, corresponding to the Fe $t_{2g}$ and $e_g$ orbitals coupled with oxygen bands, respectively. Interestingly, these features are nearly identical to un-doped $BiFeO_3$ (green lines in Fig. 2a), although the number of vacancies in the two samples is completely different. Moreover, the polarization dependence ($E//a$ and $E//c$), which is sensitive to orbital anisotropy[24] and crystal symmetry[25], is similar to $BiFeO_3$. This means that even with substantial oxygen vacancies, a change in the electronic structure of oxygen is hard to observe by this XAS measurement.

For the next step, we employed the site-selective spectroscopic technique – resonant soft X-ray scattering (RSXS). Since the oxygen vacancies in $Ca^{2+}$ doped $BiFeO_3$ are confined periodically[15], this allows the exploration of electronic configurations around an oxygen vacancy. Figure 2c shows a $\theta$–$2\theta$ scan of BCFO25 at $E$ ~ 525 eV – details of the experimental geometry are shown in Fig. 2b. It clearly shows the superstructure reflection, $q$ = (001), indicating periodically confined vacancies, which is consistent with the structural formation as shown in Fig. 1a. To investigate the site-selective (i.e., confined vacancies) spectroscopic features, energy scans at fixed $q$ were performed with two ($\sigma$ and $\pi$) incident polarizations (Fig. 2d). Note that electronic anisotropy can be resolved by controlling $\sigma$ or $\pi$ polarization in this RSXS measurement[26,27]. The measured RSXS profiles are quite unlike XAS spectra, showing the polarization dependence around the anisotropic Fe $t_{2g}$ and $e_g$ orbital bands hybridized with oxygen. In this context, the difference in the RSXS intensity profile between $\sigma$ and $\pi$ represents the anisotropic Fe $3d$ orbital state as modified by the oxygen vacancies.

Since Fe cations in BCFO are chemically correlated with the oxygen vacancies, we need to scrutinize the Fe electronic structures. Figure 3a shows the XAS spectrum for the Fe $L_{2,3}$-edges. The spectral features are almost identical to the known $Fe^{3+}$ cation feature[23,28]. This means that the Fe valence is retained as a single 3+ state. This is in agreement with atomic multiplet calculations[29] on the single valence state under $D_{4h}$ symmetry (see Methods). Moreover, this calculation can generate linear dichroism (LD = $E//a - E//c$). The calculated LD is comparable to experimental results (Fig. 3b) except for a small deviation around the in-plane orbital characters ($xy$ and $x^2$-$y^2$). This deviation, in particular the $x^2$-$y^2$ character, is more pronounced in more heavily doped system (30% doped BCFO30). These findings might be associated with the implication (i.e., Fe $3d$ orbital state modified by oxygen vacancy) of O $K$-edge RSXS measurements. In other words, the in-plane orbital characters as modified by the oxygen vacancies undergo an additional anisotropic effect beyond the tetragonal crystal symmetry.

We now consider RSXS measurements at the Fe $L_{2,3}$-edges, for exploring Fe orbital anisotropy around the oxygen vacancies. Like the observed superstructure at the O $K$-edge, we clearly see a superstructure reflection at $\boldsymbol{q}$ = (001), in addition to the second order (002) reflection (Fig. 4a inset). In the Fe $L$-edge RSXS study, we focused on the $\boldsymbol{q}$ = (002) peak of BCFO25. Figure 4a shows the Fe $L$-edge RSXS profile for $\sigma$ incident polarization. Note that the Fe profiles have been subtracted by a diffuse scattering part, e.g. fluorescence background (see Supplementary Information). In comparison with the Fe XAS spectrum, the RSXS profile is quite complicated. This complexity arises from modification of the Fe local structure by the oxygen vacancies. The elongated octahedral Fe ($D_{4v}$) coordination in doped BiFeO$_3$ can be transformed to tetrahedral ($T_d$) and square pyramidal ($C_{4v}$) symmetry via oxygen vacancies[19,30]. The resonant scattering is produced by the scattering form factor which is basically determined from the crystal symmetry. Therefore, the Fe RSXS profile in Fig. 4a is constructed by all symmetries in the BCFO. Accordingly, the current RSXS profile corresponding to both the complicated structural effects and regarding their Fe spectroscopic information causes a difficulty in exploring the Fe 3$d$ orbital state modified by the oxygen vacancy.

To overcome this difficulty, we employed polarized X-rays and the principle of Brewster's angle[31] in this measurement. Moreover, this is why we focused on the (002) peak of BCFO25 (see Supplementary Information). Note that we do not control a polarization of the out-going photon, indicating the scattered X-ray always shows both $\sigma_f$ and $\pi_f$ polarizations. Considering the principle, in here $\theta_i + \theta_f \sim 90°$ Brewster geometry, structural contribution (via $\pi_i - \pi_f$ channel) is drastically suppressed in incident $\pi_i$-polarization, while the structural contribution (via $\sigma_i - \sigma_f$ channel) is still large in incident $\sigma_i$-polarization[32]. As a consequence, we clearly observed the Fe spectroscopic behavior via the $\pi_i - \sigma_f$ channel of incident $\pi_i$-polarization at $\boldsymbol{q}$ = (002) (shown in Fig. 4b). Remarkably, there are only two pronounced features around $E$ = 706 eV and 708 eV, agreeing with the implication of the Fe $L$-edge XAS measurements, which respectively corresponds to $xy$ in $t_{2g}$ orbital bands and $x^2$-$y^2$ in $e_g$ orbital ones. This indicates that the $Fe^{3+}$ band, in particular in-plane orbital bands, becomes anisotropic around the Fermi level, revealing the role of oxygen vacancy in BCFO system.

## Discussion

Considering the Fe octahedral structure in the BCFO, the in-plane orbital characters in the crystal symmetry of the BCFO is not energetically preferred because of the $c$-axis elongation, showing the self-assembled structure as shown in Fig. 1a. Nevertheless, local in-plane Fe orbital bands on the self-assembled layers formed by the oxygen vacancy are clearly reconstructed by the hybridization with the vacancy. This reconstruction behavior is clearly observed when the structural effect is suppressed through Brewster geometry in RSXS measurement, leading to the additional anisotropic effect in the doped BCFO. Eventually, electrons hopping behavior around the Fermi level is disturbed by the additional anisotropic effect that attributes to the localized orbital bands, reinforcing insulating behavior on the BCFO.

In summary, we have experimentally demonstrated the role of oxygen vacancy which is confined into the two-dimensional self-assembled layers occurring periodically at a few nanometers interval in the Ca-doped $BiFeO_3$ films by using XAS and RSXS techniques. The central finding here is that the orbital state of $Fe^{3+}$ cation is modified via the hybridization with the oxygen vacancy, which is competing with the electronic configuration of the Fe valence band in BCFO. This gives a key idea why with increasing doping ratio the diode effect of BiFeO$_3$ becomes weak even in higher contents of the oxygen vacancy in the previous report[22]. These microscopic aspects of oxygen vacancies open a window into a new regime of energy materials, and oxides in general.

## Methods

**Sample preparation.** Using pulsed laser deposition (KrF excimer laser, $\lambda$ = 248 nm), BCFO films were grown on SrTiO$_3$ (001) substrates at 600–700 °C in 50–100 mTorr oxygen pressure. The films were cooled down at a rate of 5 °C/min with an oxygen pressure of ~ 1 atm. With increasing the $x$ ratio, practically the oxygen vacancy in BCFO film is increasing[15,18]. As varying Ca substitution ratio ($x$ = 0.075 ~ 0.30), we monitored BCFO films' superstructural form, including crystalline quality, using by X-ray diffraction with Cu $K_{\alpha 1}$ ($\lambda$ = 1.54 Å) radiation (see Fig. 1b). Aiming to manipulate the periodicity of the oxygen vacancy which is confined around interfaces of the superstructure, finally, the $x$ range was chosen to 0.20, 0.25, and 0.30 (see Supplementary Information).

**Synchrotron experiments.** The XAS spectra show white line resonances at the Fe $L_{2,3}$-edges. The spectra result from Fe $2p \rightarrow 3d$ dipole transitions, are divided roughly into the $L_3$ ($2p_{3/2}$) and $L_2$ ($2p_{1/2}$) regions. For the LD measurements via XAS, the polarization direction of the linearly polarized X-rays (98% polarized) was tuned by elliptically polarized undulator, with horizontal ($\sigma$) and vertical ($\pi$) polarizations corresponding to complete in-plane ($E//a$) and majority out-of-plane ($E//c$) polarized components, respectively (see Fig. 2b). Theses spectroscopic experiments, XAS and RSXS, were carried out at beamlines 8-2 and 13-3 of the Stanford Synchrotron Radiation Lightsource (SSRL). Note that all measurements were done by zero-electric field polarization.

**Atomic multiplet calculations.** The calculations were carried out for the configuration interaction via combination between the initial $2p^63d^5$ state and its charge transfer $2p^63d^6\underline{L}$ state under the $D_{4h}$ crystal symmetry. The used Coulomb interactions are $U_{dd}$ = 5 eV and $U_{pd}$ = 6 eV. The charge transfer energy is $\Delta$ = 2.7 eV. The crystal field ($10D_q$ = 1.6 eV) was used for this calculation. The slater integrals are with ~ 80% of the atomic values.

## Acknowledgements

We thank Dr. Kevin Hunter Stone for useful discussions. Synchrotron studies were carried out at the SSRL, a Directorate of SLAC and an Office of Science User Facility operated for the US DOE Office of Science by Stanford University. J.L. acknowledges support by the Department of Energy, Office of Basic Energy Sciences, Materials Sciences and Engineering Division, under contract DE-AC02-76SF00515. This work at KAIST was supported by the National Research Foundation (NRF) Grant funded by the Korean Government (NRF-2013S1A2A2035418, 2014R1A2A2A01005979).

## Author contributions statement

H.J., C.Y., and J.-S.L. planned the experiments. H.J., G.K., and J.-S.L. carried out the Fe $L$-edge and O $K$-edge soft x-ray diffraction experiments at SSRL. H.J., C.Y., C.K., and J.-S.L. contributed to the data interpretation and modeling discussions. J.L. and C.Y. grew the samples. H.J. and J.-S.L. wrote the manuscript with contributions from the other authors. J.-S.L. is responsible for overall project direction, planning, and management.

## Additional information

Competing financial interests: The authors declare no competing financial interests.

# Figures

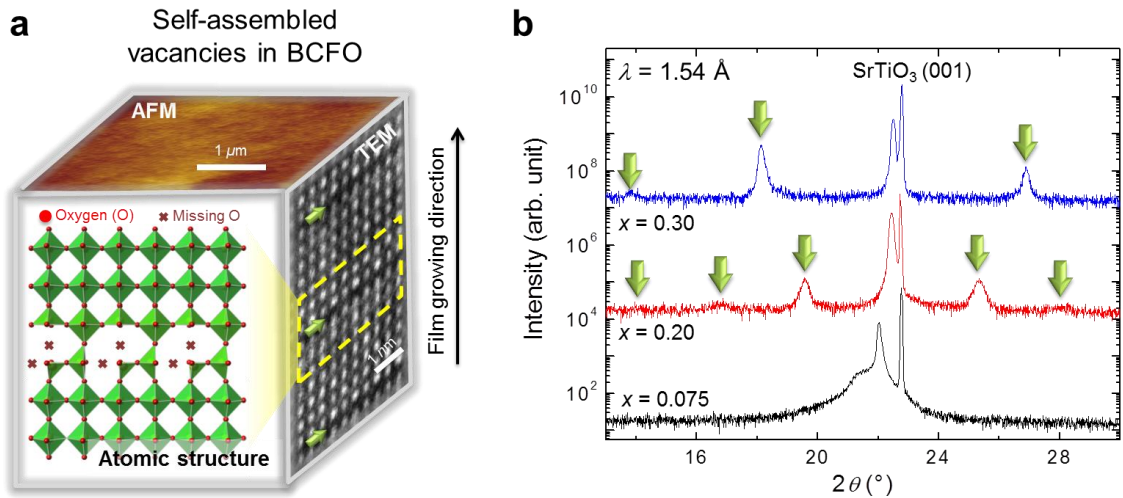

**Figure 1.** (a) Conceptual atomic structure shows a schematic picture of the self-assembled vacancies in the doped BCFO. AFM reveals the flatness of the film surface. TEM image, taken from our previous work[20] [Copyright notice. Reprinted with the permission], indicates the location oxygen vacancy with respect to the growing direction. The green arrows indicate the vacancy position. (b) Confirmation of the superstructural formation of oxygen vacancy in BCFO films using X-ray diffraction. The films ($x \geq 0.2$) show clear superstructure (indicated by green arrows) reflections. Central sharp peak corresponds to substrate $SrTiO_3$ (001) and adjacent lower angle peaks indicate BCFO film.

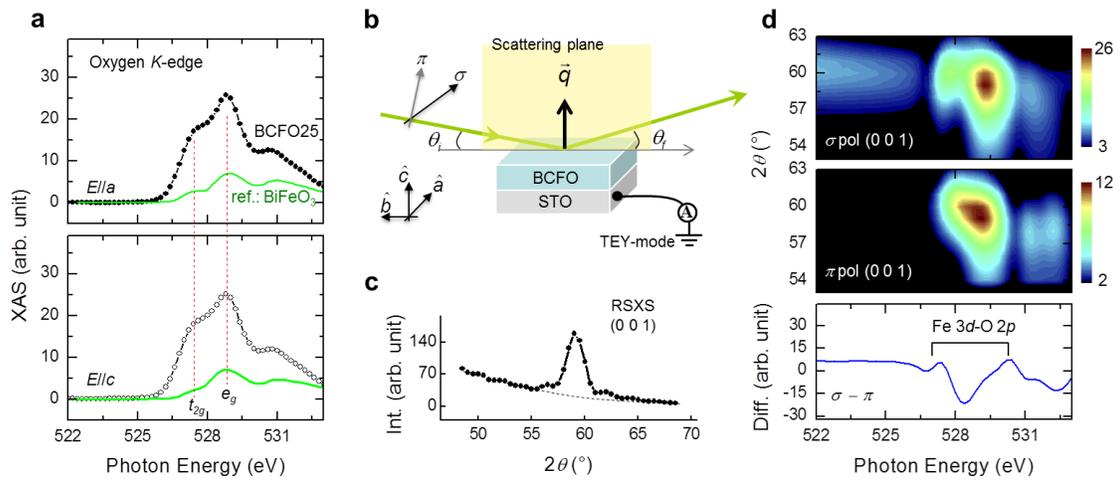

**Figure 2.** (a) O *K*-edge XAS spectra of BCFO25 with *E // a* and *E // c*. The spectra are compared to the scaled reference (bulk BiFeO$_3$) spectra. Dashed red lines denote energy positions of $t_{2g}$ and $e_g$ orbital states (b) Experimental configuration for XAS and RSXS experiments. $\theta_i$ and $\theta_f$ denote incident and scattered angles, respectively. For XAS experiment, $\theta_i$ was set by 20° – *E // a* (*E // c*) was measured in $\sigma$ ($\pi$) polarization. (c) Superstructure reflection of BCFO25, *q* = (001) at 525 eV. Dashed line is estimated by the specular background. (d) O *K*-edge RSXS profiles of (001) reflection using $\sigma$ and $\pi$ polarization, and the polarization difference of angle integrated energy profiles.

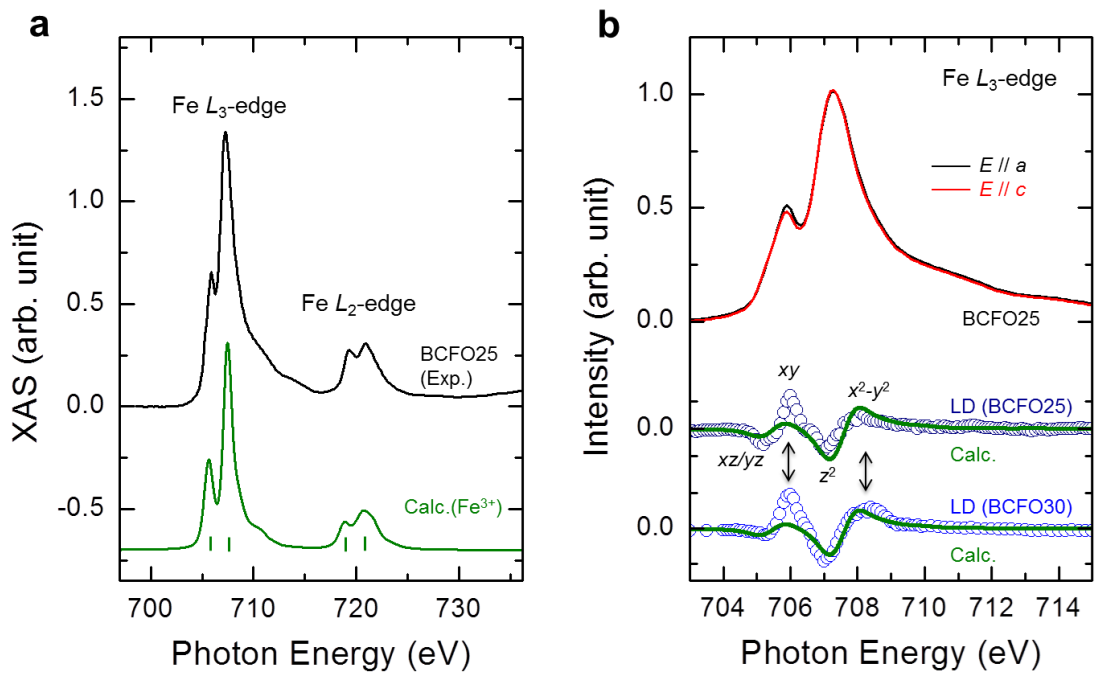

**Figure 3.** (a) XAS spectrum at Fe $L_{2,3}$-edges on BCFO25. The spectrum is compared to a calculated spectrum of $Fe^{3+}$ in $D_{4h}$ symmetry. In the calculation, the vertical bars denote the atomic states. (b) Polarization dependence of BCFO25 with $E // a$ and $E // c$. LD spectra of BCFO25 and BCFO30, which are compared with calculated LD of $Fe^{3+}$ in $D_{4h}$ symmetry. Arrows indicate a discrepancy between experiments and calculation.

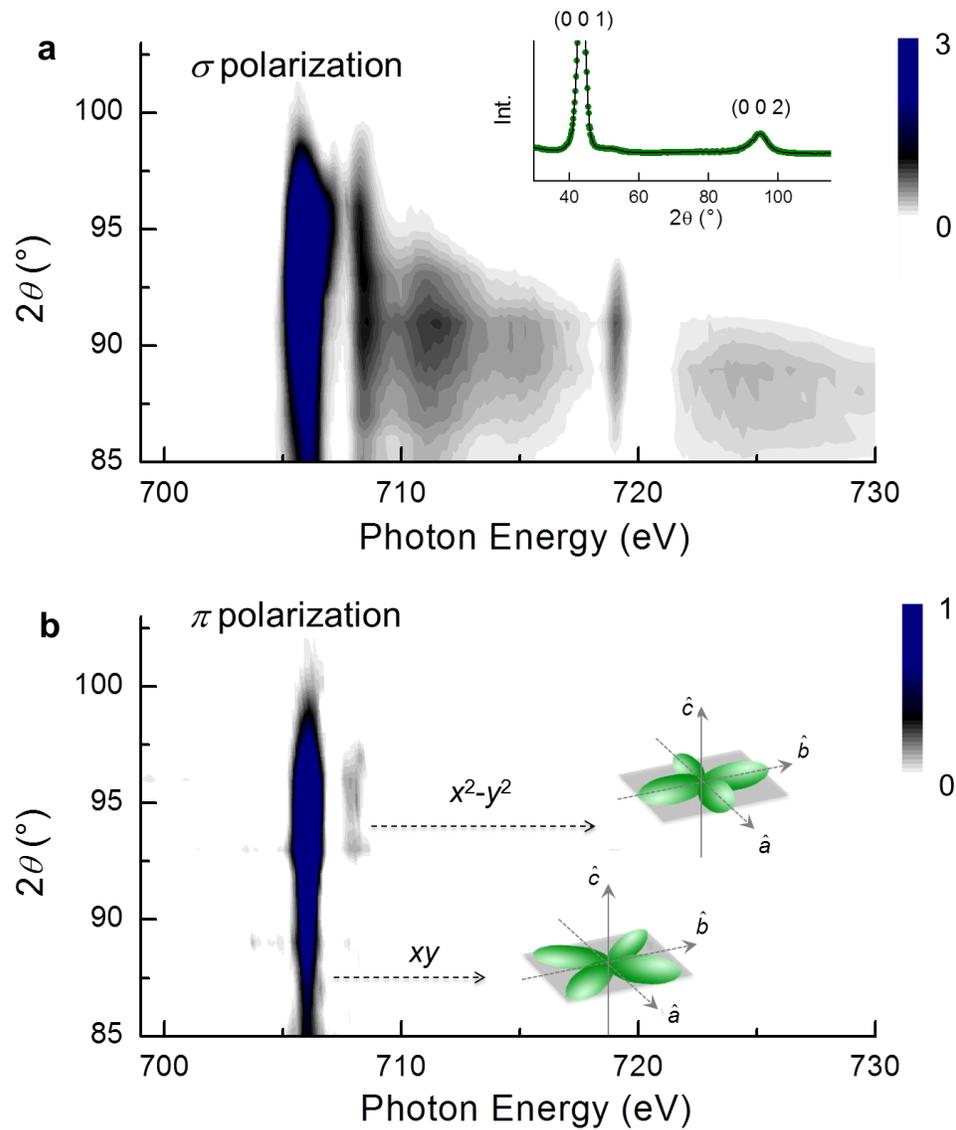

**Figure 4.** Fe $L_{2,3}$-edge RSXS profiles. All profiles are corrected by fluorescence background. (a) The profile at $\mathbf{q}$ = (002) was taken by incident $\sigma$ polarization. Inset shows full $\theta$–$2\theta$ scan. (b) The profile at $\mathbf{q}$ = (002) was taken by incident $\pi$ polarization. Two pronounced features at $E$ = 706 eV and 708 eV are corresponding to in-plane $xy$ and $x^2$-$y^2$ orbital states. Arrows indicate cartoon illustrations of those orbitals

Supplementary Information for

# Orbital Reconstruction in a Self-assembled Oxygen Vacancy Nanostructure

H. Jang[1], G. Kerr[1], J. S. Lim[2], C.-H. Yang[2], C.-C. Kao[3], and J.-S. Lee[1,*]

[1]Stanford Synchrotron Radiation Lightsource, SLAC National Accelerator Laboratory, Menlo Park, California 94025, USA

[2]Department of Physics, KAIST, Yuseong-gu, Daejeon 305-701, South Korea

[3]SLAC National Accelerator Laboratory, Menlo Park, California 94025, USA

*jslee@slac.stanford.edu

## [S. 1] Manipulation of a periodicity of oxygen vacancy

When $Ca^{2+}$ substitution rate ($x$) is larger than ~ 0.15, oxygen vacancy superstructure is formed[S1] and the periodicity empirically shows ~ $1.5/x$. When $x$ becomes 0.20, 0.25, and 0.30, the periodicity becomes 8 unit cells (u.c.), 6 u.c., and 5 u.c., respectively and corresponding superstructure reflection positions changed (Fig. S1). When $x$ = 0.25, $2\theta$ angle of second superstructure reflection (002) is near 90° and therefore the structural contributions are suppressed with $\pi$ polarized incident light (see section [S. 2] for details).

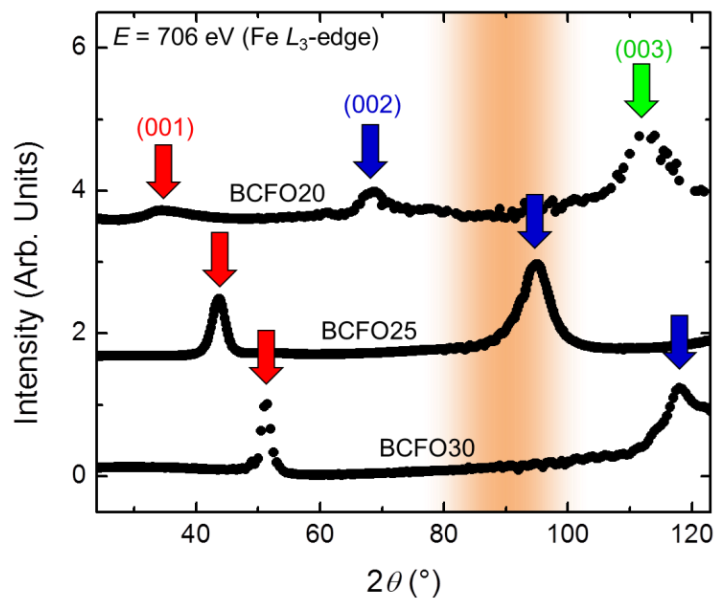

Figure S1. Manipulating the periodicity of the oxygen vacancy ordering by $Ca^{2+}$ substitution ratio. First, second, and third superstructure reflections are indicated by red, blue, and green colored arrows, respectively. The intensities are divided by $Q^4$ and shifted for clarity, where $Q$ is the absolute value of momentum transfer in unit of Å$^{-1}$. Orange area near $2\theta$ = 90 ° indicates the vicinity of Brewster's angle.

# [S. 2] Brewster's angle and suppression of structural signal at $2\theta = 90°$ and selection of (002) reflection of BCFO25

When a light strikes a surface, the amount of reflection and transmission is determined by refractive index of the target material ($n$), incident angle ($\theta_i$), and polarization rate of the light. Geometry of incident and scattered light is shown in Fig. S2a. At a certain incident angle, no $\pi$ polarized light is reflected. The angle is called as Brewster's angle ($\theta_B = 90° - \theta_i$) and determined by $\tan\theta_B = n/n_0$, where $n_0$ is a refractive index of vacuum[S2]. In x-ray region, $n$ is very close to 1, and therefore $\theta_B$ is nearly identical to 45° (Fig. S2b).

In general, intensity of Thomson scattering which is from structural contribution is determined by $(\varepsilon_i \cdot \varepsilon_f)$ where $\varepsilon_i$ ($\varepsilon_f$) is polarization of incident (scattered) light[S3]. Since we only control a polarization of the incident x-ray polarization, the scattered photon's polarization always shows both $\sigma_f$ and $\pi_f$ polarizations. In this sense, when we use the $\pi$ as the $\varepsilon_i$, the term, $\pi_i \cdot \sigma_f$, becomes zero. This is because an angle between two polarizations' directions is orthogonal. Therefore, the structural contribution is determined solely by $\pi_i \cdot \pi_f$ in the $\pi$ channel. When $2\theta = \theta_i + \theta_f = 90°$ (Fig. S2b), $\pi_i \cdot \pi_f$ also becomes zero. Consequently, with $\pi$ polarized light, there is little structural contribution from either reflection or diffraction near $2\theta = 90°$ (e.g. BCFO25 (002) reflection as shown in Fig. S1). Furthermore, we need to consider an addition polarization term in the resonant x-ray scattering, i.e., $(\varepsilon_i \times \varepsilon_f)$[S3], showing the 2 × 2 polarization matrix. Diagonal terms in the matrix are proportional to isotropic form factor (i.e., structural factor), which is similar effect in $(\varepsilon_i \cdot \varepsilon_f)$. Off-diagonal terms in the matrix are sensitive to detect anisotropic contributions such as spin, orbital, and charge disproportion.

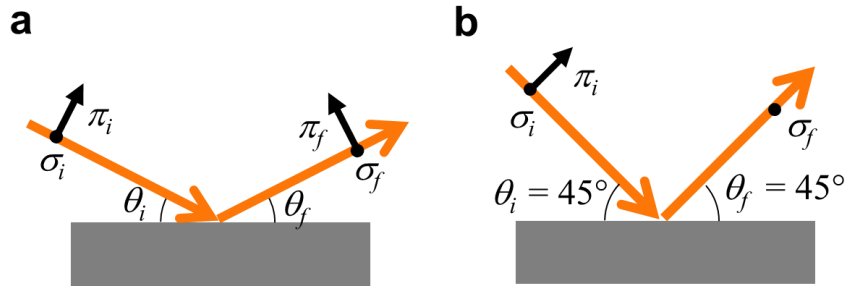

Figure S2. (a) Simple description of incident and scattered x-ray geometry with incident ($\theta_i$) and scattered angle ($\theta_f$) with linear polarization ($\sigma,\pi$). (b) At Brewster's angle position ($\theta_i = \theta_f = 45°$), no structural contribution from $\pi$ polarized light.

## [S. 3] Subtraction of diffuse scattering

Even in diffraction condition, unexpected signals can be detected as well as pure diffraction signal. One of them is fluorescence background. It is photon energy dependent but there is little angle dependence. Therefore, if we subtract a signal from out of diffraction condition, we can remove fluorescence background.

Fig. S3a shows rocking ($\theta$) scan of BCFO25 (002) reflection. Figure S3b and S3c show RSXS profiles at both on diffraction condition (red colored arrow in Fig. S3a) and diffuse condition conditions (blue colored arrow in Fig. S3a) of (002) in incident $\sigma$ polarization and $\pi$ polarization, respectively. $\sigma_i$ RSXS profile contains complicated structural modification by oxygen vacancy (Fig. S3b). In $\pi_i$, structural contribution is suppressed, and therefore nearly pure electronic signal can be obtained (Fig. 4b in main article). (001) reflection which is away from Brewster's angle show complicated structural contribution in both $\sigma_i$ and $\pi_i$ cases (Fig. S4).

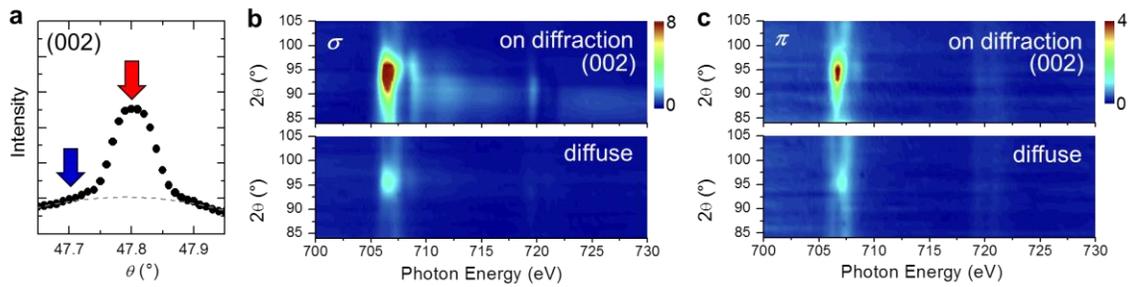

Figure S3. (a) Red and blue colored arrow indicate the position of on diffraction and diffuse conditions of (002) reflection. RSXS profiles on diffraction condition and diffuse condition of (b) (002) with $\sigma_i$ and (c) (002) with $\pi_i$. Complicated structural contribution is suppressed with $\pi_i$ in contrast to structural contribution dominant $\sigma_i$ case.

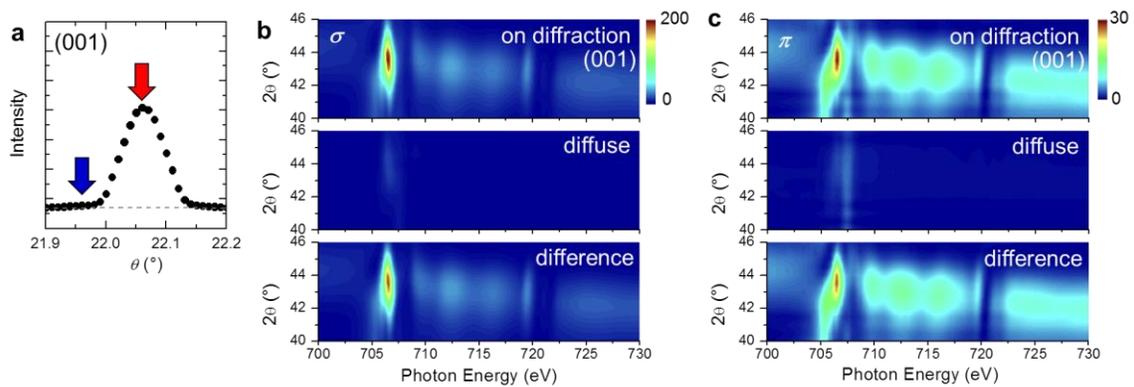

Figure S4. (a) Red and blue colored arrow indicate the position of on diffraction and diffuse conditions of (001) reflection. RSXS profiles at on diffraction condition, diffuse condition, and their difference of (b) (001) with $\sigma_i$ and (c) (001) with $\pi_i$